\documentclass[twocolumn]{aastex631}

\begin{document}

\title{
Exploring Substructure of the Near-Surface Shear Layer of the Sun
}

\correspondingauthor{M. Cristina Rabello Soares}
\email{cristina.rabello.soares@stanford.edu}

\author[0000-0003-0172-3713]{M. Cristina Rabello Soares}
\affiliation{W. W. Hansen Experimental Physics Laboratory, Stanford University, Stanford, CA, 94305-4085, USA}

\author[0000-0002-6163-3472]{Sarbani Basu}
\affiliation{Department of Astronomy, Yale University, PO Box 208101, New Haven, CT 06520-8101, USA}

\author[0000-0002-0910-459X]{Richard S. Bogart}
\affiliation{W. W. Hansen Experimental Physics Laboratory, Stanford University, Stanford, CA, 94305-4085, USA}

\begin{abstract}

The gradient of rotation in the near-surface shear layer (NSSL) of the Sun provides valuable insights into the dynamics associated with the solar activity cycle and the dynamo. 
Results obtained with global oscillation mode-splittings lack resolution near the surface, prompting the use of the local helioseismic ring-diagram method. 
While the Helioseismic and Magnetic Imager ring-analysis pipeline has been used previously for analyzing this layer, default pipeline parameters limit the accuracy of the near-surface gradients. 
To address these challenges, we fitted the flow parameters to power spectra averaged over one-year periods at each location, followed by additional averaging over 12 years.
We find that the NSSL can be divided into three fairly distinct regions: a deeper, larger region with small shear, steepening towards the surface; a narrow middle layer with a strong shear, with a gradient approximately three times larger; and a layer very close to the surface, where the logarithmic gradient is close to zero but becomes steeper again towards the surface. The middle layer appears to be centered at 3 Mm, but the poor resolution in these layers implies that it is potentially located closer to the surface, around 1.5 Mm deep. 
While our analysis primarily focused on regions along the central meridian, we also investigated systematic errors at longitudes off the center. The east-west antisymmetric component of the gradient reveals a layer of substantial differences between east and west longitude around at 1.7 Mm, and the amplitude of the differences increases with longitude.

\end{abstract}


\section{Introduction} \label{sec:intro}

Helioseismology has significantly advanced our understanding of the Sun's internal rotation, providing crucial insights such as the identification of two regions of radial shear in the rotation rate. One of these regions, commonly referred to as the 'tachocline' \citep{spiegel1992}, is situated near the base of the convection zone; this region near the interface between the convective and radiative zones is believed to serve as a key location for magnetic field storage \citep{parker1975}. 
The other region, known as the near-surface shear layer (NSSL), is located close to the solar surface, extending down to a depth of about 35 Mm, i.e., approximately 5 percent of the solar radius \citep{rhodes1990}. 
Ongoing research increasingly delves into the role of the NSSL in solar dynamos, with studies examining its potential impact \citep[e.g.,][and references within]{karak2016}.

Various studies have explored the determination of the rotational gradient within the NSSL, employing diverse data and methodologies. However, a consensus on its characteristics remains elusive. 
\citet{corbard2002} and \citet{barekat2014, barekat2016}, used only {\sl f-}mode global helioseismic data and thus limited the range of  spherical harmonic degree analyzed, which confined their investigations to a radius between approximately 0.986 and 0.994 R$_\odot$ (depths between 10 and 4 Mm). They observed a logarithmic radial gradient, $\partial \ln \Omega/\partial \ln r$, very close to $-1$.
\citet{antia2022} employed a 2D regularized least-squares (RLS) technique on global {\sl f-} and {\sl p-}modes, enabling analysis of a broader range in spherical harmonic degree and enabling measurements over a wider range of radius. Their approach, employing B-splines and analytic derivatives, reduces noise in estimating the rotational gradient. They reported a logarithmic radial gradient close to $-1$ at 0.99 R$_\odot$ (a depth of 7 Mm), decreasing in amplitude to approximately $-0.2$ at 0.95 R$_\odot$ (35 Mm).
The local helioseismic analysis technique, known as ring-diagram analysis \citep{hill1988}, has also been used to estimate the rotational gradient near the surface. This technique is based on fitting three-dimensional power spectra of tracked regions on the Sun. Because the analysis provides data on modes of higher spherical harmonic degrees, it gets better results closer to the surface. 
\citet{basu_etal1999} conducted a ring-diagram analysis on patches of diameter 15 heliographic degrees, revealing a more pronounced slope in zonal flow at depths less than 4 Mm. \citet{zaatri2009}, in their ring-diagram analysis of 15-degree patches, observed a relatively constant radial gradient between 0.980 R$_\odot$ and approximately 0.991 R$_\odot$ (14 and 6 Mm deep), specifically below 40$^\circ$ latitude. In this range, the logarithmic gradient remains approximately $-0.76$, as illustrated in their Figure 1. Notably, above 0.992 R$_\odot$ ($6$ Mm), the logarithmic gradient becomes increasingly negative, extending at least up to 0.996 R$_\odot$ (3 Mm).
\citet{komm2022}, using standard ring-diagram results from GONG and HMI pipelines, reported a logarithmic gradient equal to $-1$ at 0.990 R$_\odot$, consistent with previous studies. Similar to \citet{zaatri2009}, he observed a steeper gradient closer to the surface, with the steepest slope at 0.998 R$_\odot$ (1.4 Mm).

\citet{barekat2014, barekat2016} observed a very slight increase in the gradient with latitude in the range $r =$ 0.990 $\pm$ 0.004 R$_\odot$, similar to \citet{antia2022} at 0.95, 0.97, and 0.99 R$_\odot$ (35, 21, and 7 Mm). In contrast, \citet{corbard2002} and \citet{zaatri2009} reported a more pronounced variation with latitude above 40$^\circ$, where the value of the dimensionless gradient tends towards zero at 60$^\circ$. \citet{reiter2020} extended their analysis to spherical harmonic degree as high as 1000, revealing a substantial variation in the logarithmic gradient with latitude very close to the solar surface, increasing in steepness from approximately $-1.5$ at the equator to $-2.8$ at 28.1$^\circ$ latitude, then approaching zero at 70$^\circ$ latitude.
These nuanced and evolving findings highlight the complex challenges in characterizing the NSSL. As research advances, the importance of establishing the NSSL's behavior becomes increasingly apparent, particularly for understanding its pivotal role in shaping solar dynamics and its connection with phenomena such as sunspots and the broader solar magnetic cycle.

The variation of the gradient over time during one or more solar cycles has been consistently reported to be small, with $\partial \ln \Omega/\partial \ln r$ within $\pm$ 0.1 or lower, as demonstrated in previous studies \citep{barekat2016,antia2022,komm2022}.

In this study, we conducted a comprehensive analysis to determine the averaged logarithmic radial gradient 
of the solar rotation rate, with a specific focus on the NSSL. Employing the local helioseismology technique of ring-diagram analysis, we analyzed 12 years of HMI data, equivalent to approximately one solar cycle. In contrast to \citet{komm2022}, our approach involved using multiple inversion methods, as well as multiple inversions using many tradeoff parameters, allowing for a more precise analysis, with reduced noise introduced by error correlation and a more accurate determination of the depth structure of the gradient than those provided by the single pipeline inversion. The next section describes the data used and the steps taken to estimate the non-dimensional rotation radial gradient. We describe our results in Section~3, and include a comparison with simulations. The results are summarized and discussed in section~4.

\section{Methodology and Data} \label{sec:data}

Our approach involved the application of ring-diagram analysis to 15-degree and 30-degree diameter regions observed by the Helioseismic and Magnetic Imager (HMI) \citep{scherrer2012} from May 2010 through April 2022.
While the HMI pipeline is an extremely useful and widely used tool, providing estimates of horizontal flows as a function of solar radius near the surface, we have identified limitations associated with the default parameters that impact the accuracy of near-surface gradients. A significant challenge in our analysis was the presence of oscillatory patterns induced by error correlations, a phenomenon highlighted by \citet{howe_thompson}. Random noise in observed mode frequencies can introduce spurious oscillatory behavior in the inverted depth structure of flows.
These oscillatory patterns are noticeable and pervasive in the inversion results, and are exacerbated when estimating the radial gradient of rotation.
To address this issue, we computed flow parameters by fitting power spectra averaged over one-year intervals at each location that we investigate. This is not the usual analysis done in the HMI ring-analysis pipeline.

The HMI pipeline tracks regions in Dopplergrams of diameter 15 and 30 heliographic degrees, spaced at 7$^\circ$.5 and 15$^\circ$ in latitude and longitude respectively.
The tracking durations for these regions are 28.8 hours and 57.6 hours respectively, approximately corresponding to the average displacement due to solar rotation at the Carrington rate being equal to their diameter. Note that the regions are tracked using the module `{\tt mtrack}' of the HMI pipeline \citep[]{Bogart_pipeline}.

Ring diagrams are essentially three-dimensional power spectra in the frequency-wavenumber space (one dimension for temporal frequency, two for spatial wavenumbers in the zonal and meridional directions). All spectra for each Stonyhurst location within a given Carrington rotation were averaged together (24 15-degree tile spectra and 12 30-degree tile spectra per location). Subsequently, we generated twelve annual spectra by averaging the rotation-averaged spectra for all Carrington rotations falling wholly or partly within the year.
Typically, this involved averaging 14 spectra, occasionally 15.
It may be noted that rotations at the endpoints were often incorporated into two of the annual averages.
Our dataset starts with observations taken on May 1st 2010, corresponding to Carrington Time 2096:240.

\begin{figure}
	\includegraphics[width=\columnwidth]{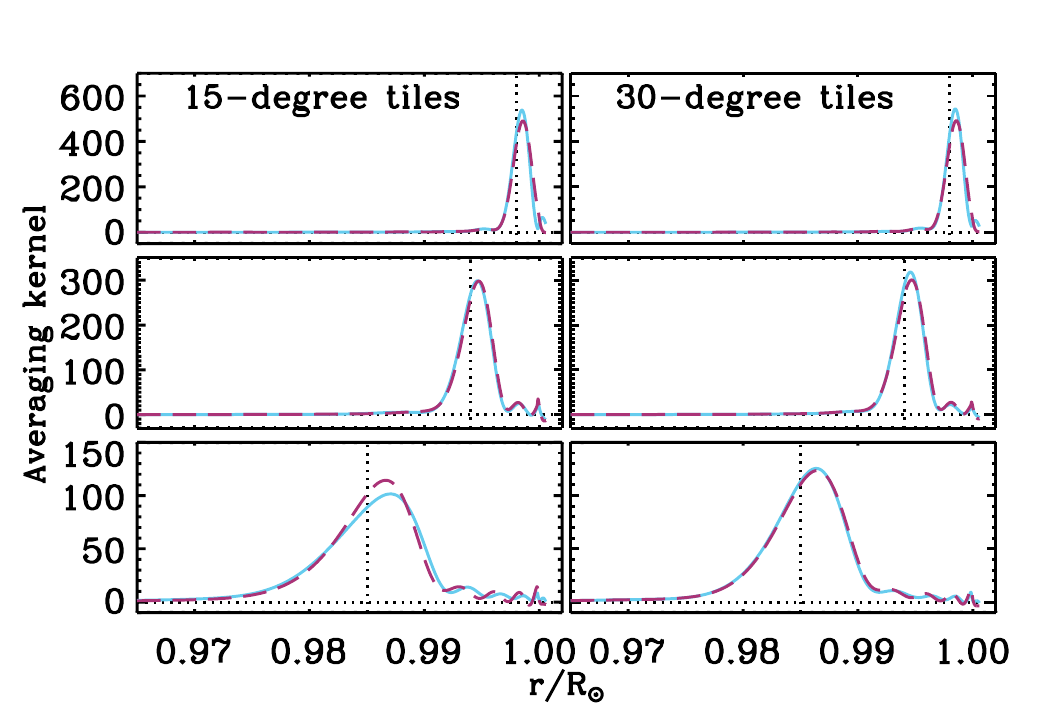}
    \includegraphics[width=\columnwidth]{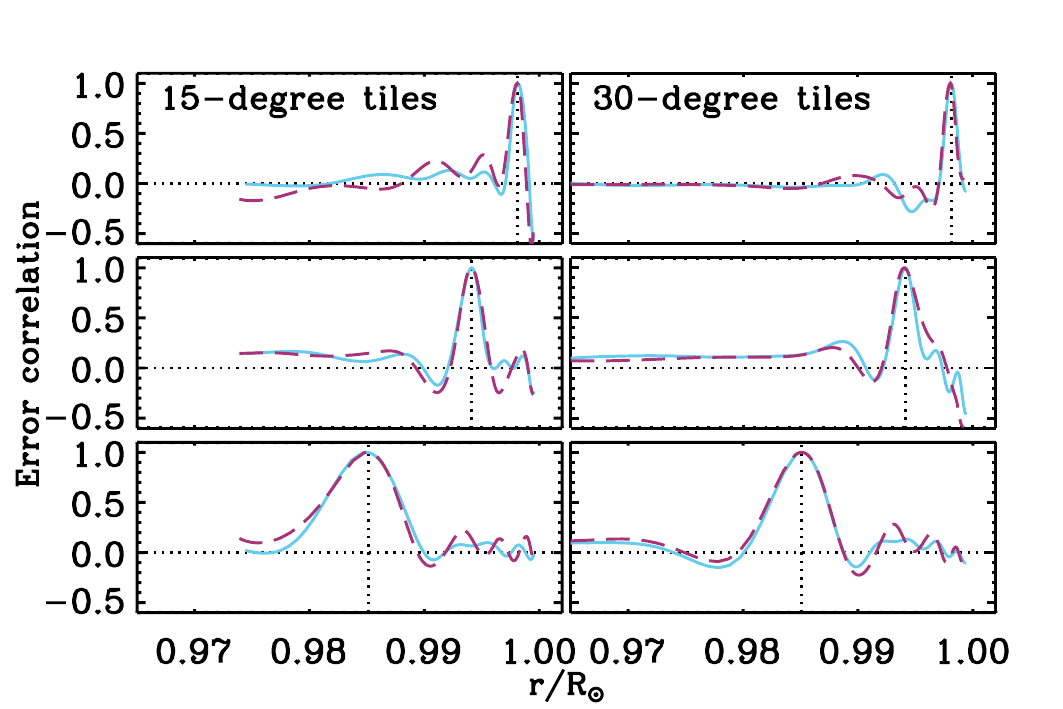}
    \caption{
    Averaging kernels (top) and error correlations (bottom) are depicted for the 15-degree tiles (left) and 30-degree tiles (right) at two latitudes for three target depths (0.985, 0.994, 0.998 R$_\odot$). The two latitudes are the equator (solid blue lines) and 60$^\circ$ north (dashed red lines). The normalization of the averaging kernel ensures that its integral is set to one.
    }
    \label{fig:avk}
\end{figure}

The annually averaged power spectra at each latitude on the central meridian were fitted using the method outlined by \citet[]{basu_antia1999} using module `{\tt rdfitc}' of the HMI processing pipeline \citep[]{Bogart_pipeline}. The resulting fitted $U_x$ flow parameter, characterizing the zonal component of the transverse flow field, was averaged over a span of 12 years for each individual mode ($n$, $\nu$), reducing the impact of noise in our analysis. We also fitted spectra at multiple Stonyhurst longitudes along the equator.
We applied two inversion techniques, Optimized Local Averaging \citep[OLA, ][module `{\tt rdvinv}' of the HMI processing pipeline]{backus1968} and the Regularized Least Squares (RLS) implementation of \citet{basu_etal1999}
to obtain the zonal flow variation with depth. Given the complementary nature of the OLA and RLS inversions \citep[see, e.g.,][]{sekii}, we can be more confident of the results if the two inversions agree. Thus by comparing the OLA and RLS results, we are able to achieve more reliable solutions for the rotation gradient. Instances of disagreement prompt a more cautious interpretation of the inversion results. 
Figure~\ref{fig:avk} presents the averaging kernel and the error correlation for 15-degree and 30-degree tiles at three target depths and two solar latitudes on the central meridian.

%%%%%%%%%%%%%%%%%%%%%%
%%%%%%%%%%%%%%%%%%%%%%

The inversion coefficients depend on the specific mode set, comprising of sets of 
radial order ($n$), spherical-harmonic degree ($l$), temporal frequency ($\nu$)
%$n$, $l$, $\nu$ 
and the fitted $U_x$ parameter (and its associated uncertainties) that are determined to be good fits.
To enhance the robustness of our inversion analysis, we generated 1,000 distinct mode sets by introducing diverse random realizations of Gaussian noise to the fitted $U_x$ flow,
where the variance is the square of the $U_x$ fitted uncertainty. Employing a comprehensive approach involving trade-off diagrams, the quality of the averaging kernels, and amount of error correlation, we meticulously selected the optimal error parameter for each tile size in both OLA and RLS methods.

To account for potential variations arising from our choice of inversion parameters, we computed the inversion coefficients for error parameters encompassing a range of approximately $\pm$20\% around our chosen value. 
This deliberate consideration of variations enabled a comprehensive exploration of the solution space.

\section{Results}

\subsection{The general structure of the NSSL}

\begin{figure*}
	\includegraphics{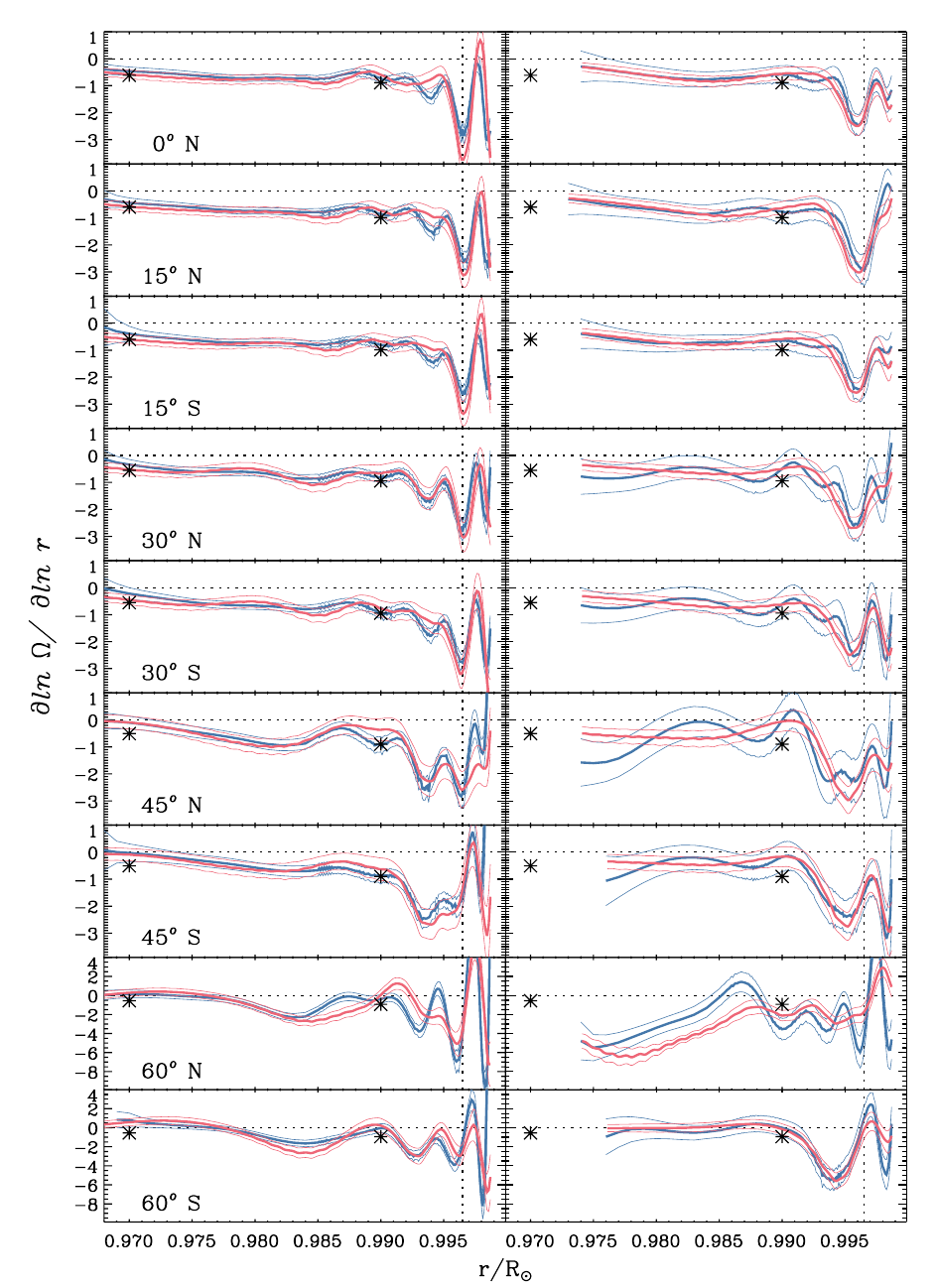}
    \caption{
    Rotational gradient obtained using OLA (blue) and RLS (red) inversion techniques applied to the 30-degree (left) and 15-degree (right) tiles at different latitudes. The vertical dotted lines are at 0.9965 R$_\odot$ (2.4 Mm deep). Global helioseismology results from \citet[]{antia2022} for HMI are symbolized by stars at 0.97 and 0.99 R$_\odot$ (21 and 7 Mm).
    }
    \label{fig:results15_30_96}
\end{figure*}

Figure~\ref{fig:results15_30_96} shows our results 
for the dimensionless radial gradient of solar rotation
($\partial \ln \Omega/\partial \ln r$)
at various latitudes along the central meridian. Solar rotation ($\Omega$) was determined by adding the the Carrington rotation velocity used to track the tiles to the velocity of the flows in the zonal direction. The results in the figure are the mean and standard deviation derived from the ensemble of 1,000 mode sets corresponding to each error parameter. This approach ensures the reliability and accuracy of our inversion results, as well as a correct estimate of the uncertainties, providing a robust foundation for the conclusions drawn in this study.

\begin{figure}
 \includegraphics[width=\columnwidth]{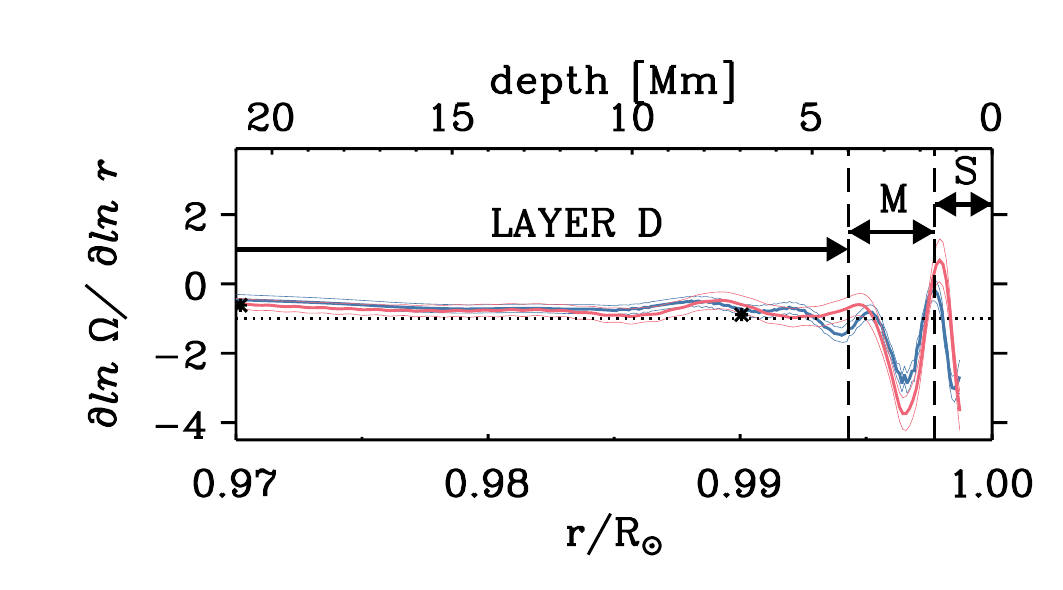}
    \caption{
    We identified three distinct regions within the NSSL: a deeper layer (D), a middle layer (M), and one closest to the surface (S).
    Rotational gradient obtained using OLA and RLS for the 30-degree tiles at the equator are shown in blue and red, respectively. Global helioseismology results from \citet{antia2022} for HMI are symbolized by stars at 0.97 and 0.99 R$_\odot$.
    The horizontal dotted line corresponds to a dimensionless gradient of $-1$.
    }
    \label{fig:layers}
\end{figure}

Figure~\ref{fig:results15_30_96} reveals that the NSSL is not a homogeneous shear layer but exhibits significant gradient variations, particularly closer to the solar surface. 
This prompt us to divide the NSSL into three specific regions, as shown in Figure~\ref{fig:layers}.            
Layer D, the deepest and thickest region, extends from 0.994 R$_\odot$ (4 Mm deep) to the edge of our inversions. 
The middle layer, Layer M, has a pronounced steepening centered around 0.9965 R$_\odot$ (2.4 Mm deep) --- indicated by the vertical dotted line in Fig.~\ref{fig:results15_30_96}. Lastly, Layer S, located very close to the surface, begins approximately at 0.9977 R$_\odot$ (1.6 Mm deep).

%%%%%%%%%%%%%%%%%%%%%%
%%%%%%%%%%%%%%%%%%%%%%

%%%%%%%%%%%%%%%%%%%%
%%%%%%%%%%%%%%%%%%%%%%

\begin{figure}
	\includegraphics[width=\columnwidth]{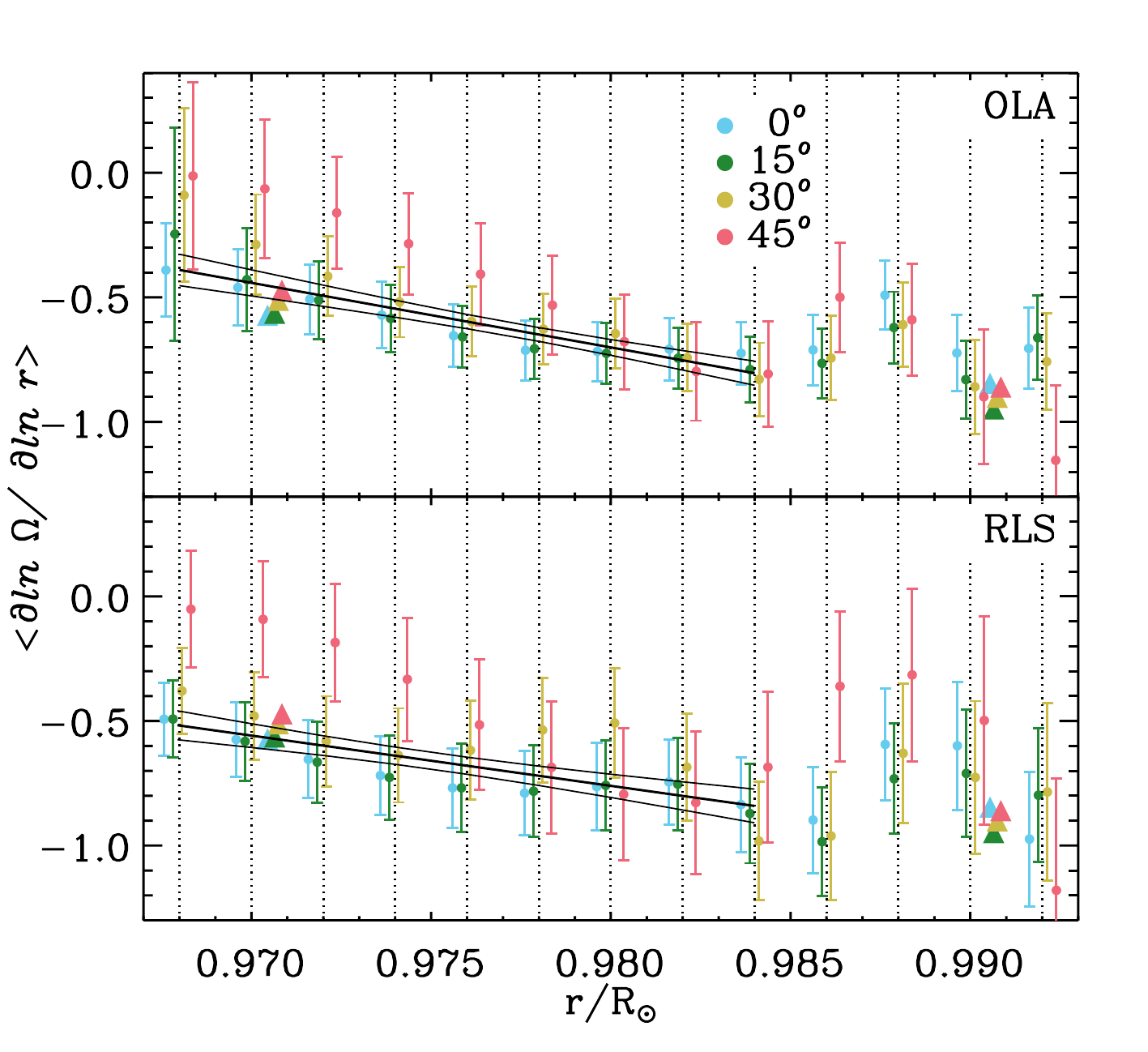}
    \caption{
    The North-South symmetric component of the logarithmic gradient of rotation at various latitudes, each represented by a distinct color. The gradients are shown at selected radii ranging from 0.97 to 0.99 R$_\odot$ with 0.002 R$_\odot$ intervals, given by the vertical dotted lines. Points are slightly displaced horizontally  for better visualization. The top panel shows results obtained using OLA, while the bottom panel shows results for RLS inversions. The triangles, slightly displaced for clarity, are the global helioseismology results from \citet[]{antia2022} at 0.97 and 0.99 R$_\odot$. The black lines represent the linear fit and its confidence level, obtained by fitting between 0.968 and 0.985 R$_\odot$ and latitudes $\leq~30^\circ$.
    }
    \label{fig:gradient985_radius}
\end{figure}

Figure~\ref{fig:gradient985_radius} shows the North-South symmetric component rotational gradient, calculated as the weighted average of the northern and southern components, denoted as (N+S)/2, in layer D. The error bars correspond to the standard deviation of the weighted mean derived from Monte Carlo simulations. Notably, a consistent linear decline with increasing radius is discernible within the depth range of $22$ to $11$ Mm (0.968 --- 0.985 $R_\odot$), where the linear fitting indicates a decrease in the gradient of 0.41$\pm$0.08 and 0.32$\pm$0.09 for OLA and RLS, respectively, between these two depths. At shallower layers of this depth interval, the gradient stabilizes around $-0.75$, persisting until 6 Mm.
The weighted average and its standard deviation are $-0.7\pm0.2$ and $-0.8\pm0.2$ for OLA and RLS respectively, between 11 and 6 Mm deep and latitudes $\leq~30^\circ$.
The results for the 15-degree tiles are similar to those of the 30-degree tiles.
A comparison with global helioseismology results, obtained by \citet[]{antia2022} and depicted in Figure~\ref{fig:gradient985_radius} with black asterisks, reveals a good agreement within the uncertainty margins.
This observation underscores the robustness of our findings.
Examining the antisymmetric component of the gradient, (N-S)/2, we find that the values fall within the range of uncertainty, suggesting a null or negligible deviation from zero, except at latitudes $60^\circ$ and higher.

%%%%%%%%%%%%%%%%%%%%%%
%%%%%%%%%%%%%%%%%%%%%%

\begin{figure*}
        \includegraphics[width=\textwidth]{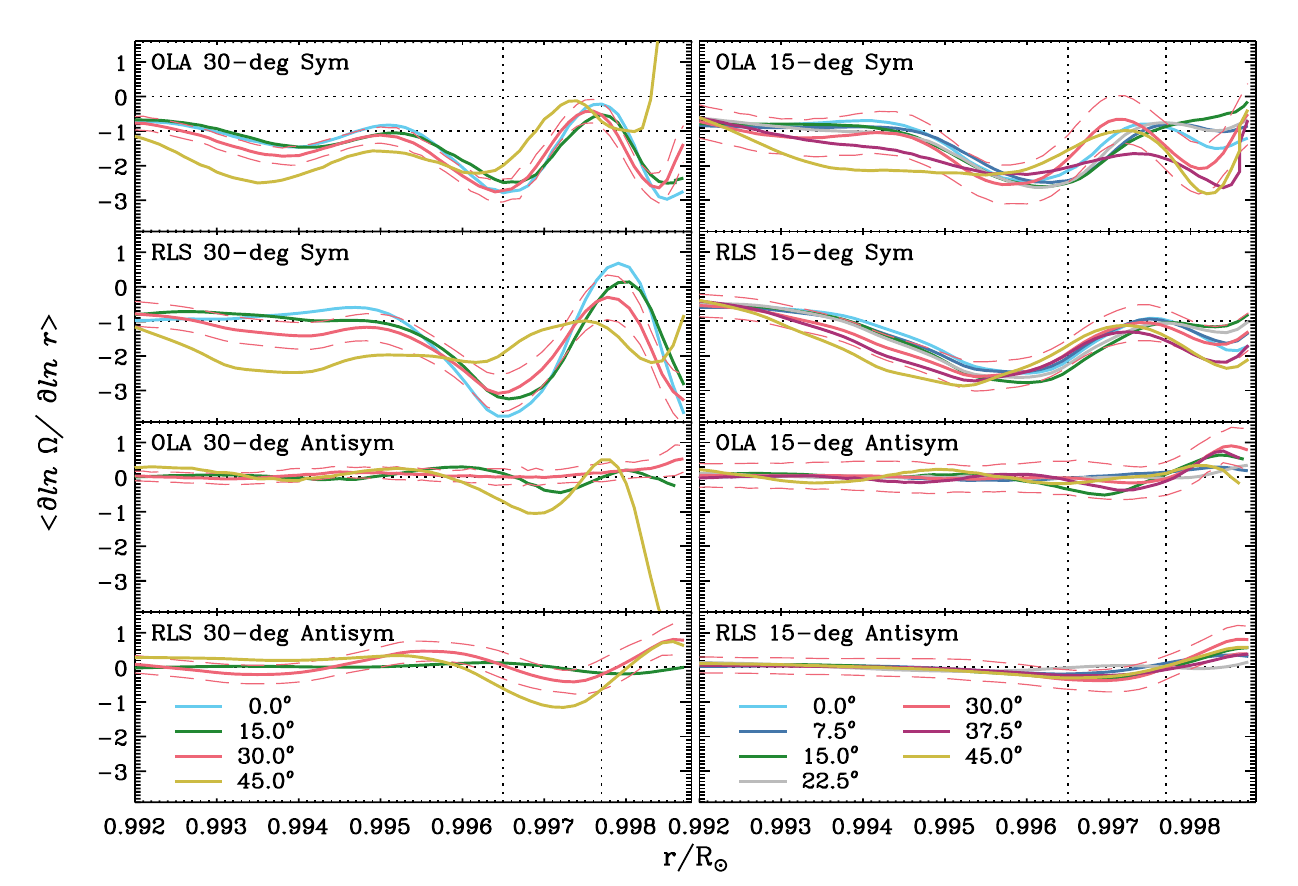}
    \caption{
    North-South symmetric and antisymmetric components of the rotational gradient close to the surface obtained using OLA and RLS inversion techniques applied to the 30-degree (left) and 15-degree (right) tiles at different latitudes. Since the uncertainties are similar, only those for 30$^\circ$ latitude are shown as dashed lines. In each panel, one vertical line indicates the location of the gradient minimum for the 30-degree tiles, the second marks the position where the gradient becomes larger than $-1$ after the dip, and correspond to depths of 2.44 Mm and 1.6 Mm, respectively.
    }
    \label{fig:sym_deep}
\end{figure*}

As one approaches the solar surface from deeper layers, a noticeable steepening of the dimensionless gradient is observed; this agrees 
with findings from other studies \citep{basu_etal1999,zaatri2009,komm2022}.
This trend begins at approximately 4 Mm depth below the solar surface (i.e., radius of 0.994 $R_\odot$), establishing the beginning of layer M, and the dimensionless gradient reaches its minimum at 3 Mm depth (radius of 0.996 $R_\odot$), as shown in Figure~\ref{fig:sym_deep}, across all latitudes up to 45$^\circ$. At shallower depths, the gradient flattens, approaching zero at all latitudes, marking the end of layer M.
Additionally, 
as is clearly visible in Fig.~\ref{fig:results15_30_96}, there is also another local minimum in the logarithmic gradient, localized
around 0.994 R$_\odot$ for latitudes up to and including $45^\circ$, but this is only seen in the OLA results; this suggests that this could be an artifact of the analysis. 
For the 30-degree tiles, both OLA and RLS results show an average minimum location of (2.44$\pm$0.07) Mm depth for latitudes below 60$^\circ$ (excluding 45$^\circ$S). 
For the 15-degree tiles, the minimum is located slightly deeper at (2.7$\pm$0.1) Mm for latitudes below 30$^\circ$ and even deeper, at (3.1$\pm$0.2) Mm, for latitudes below 60$^\circ$.  
The variation in depth between the 15-degree and 30-degree tiles can be attributed to a more precise determination of the flow parameter fitted using the 30-degree power spectra and also the larger number of modes fitted, as indicated by the smoother inversion results in the figure.

%%%%%%%%%%%%%%%%%%%%%%
%%%%%%%%%%%%%%%%%%%%%%

It should be noted that while our results obtained with the 15-degree and 30-degree tiles are generally consistent with each other, they sample different areas of the Sun. For instance, a 15-degree tile centered at 45$^\circ$ latitude spans from 37.5$^\circ$ to 52.5$^\circ$ latitude, while a 30-degree tile centered at the same latitude extends from $30^\circ$ to $60^\circ$ latitude. It should be noted that not only does each tile size differ in extension, but tiles centered off the equator are also effectively sampling regions centered at latitudes slightly lower than their nominal centers. This is simply due to the larger area at lower latitudes of a circle on a sphere centered off the equator.
Consequently, more weight is assigned to the lower latitudes inside a given tile, making the effective latitude of the tile lower than its central (nominal) latitude. The effect is small, however, amounting to displacements of about 1$^\circ$.25 for 30-degree tiles centered at latitude 60$^\circ$ and about 0$^\circ$.3 for 15-degree tiles centered at the same latitude.

The average value of the dimensionless gradient at the minimum is  around $-2.7\pm$0.1 and $-3.2\pm$0.4 for OLA and RLS inversions respectively, for latitudes below 60$^\circ$ for the 30-degree tiles; we excluded results at 45$^\circ$S because those results do not have the same pattern as the others. Similar results are obtained with the 15-degree tiles, yielding a gradient value of $-2.6\pm$0.2 for both inversion methods.

To quantify the width of this region of steep rotational gradient (within layer M), we calculate the Full Width at Half Maximum (FWHM) assuming the rotational gradient at 0.990 $R_\odot$ as the background level. For the 30-degree tiles, the average FWHM is (1.0$\pm$0.1) Mm for latitudes up to 30 degrees. The RLS results have a 10\% broader FWHM compared with the OLA results. In the case of the 15-degree tiles, the FWHM is 50\% broader than that of the 30-degree tiles, with OLA results averaging (1.3 $\pm$ 0.1) Mm and RLS (1.7$\pm$0.2) Mm for latitudes up to 45$^\circ$. The parameters of the minimum are detailed in Table~\ref{tab:minimum}.

As shown in Fig.~\ref{fig:sym_deep}, the steepening of the logarithmic gradient starts at approximately a depth of 4 Mm, reaches its minimum at 3 Mm, and is succeeded by a rapid increase, reaching approximately $-1$ for the 15-degree tiles and even larger for the 30-degree tiles --- approximately $-0.5$ using OLA and zero for RLS --- at around 1.6 Mm (0.9977 R$_\odot$). This is what we call layer M.
In contrast, Figure 1 in \citet{zaatri2009} shows the steepening beginning at about a depth of 5.5 Mm (0.992 R$_\odot$) and extends to at least 2.8 Mm (0.996 R$_\odot$), which is the minimum depth they sampled. 
In the study by \citet{komm2022}, the onset of steepening is reported at a depth of 10 Mm (0.986 R$_\odot$), reaching its steepest value at 1.4 Mm (0.998 R$_\odot$), a much shallower layer than what we see in our analysis.
We believe that this discrepancy is a result of the better resolution of the inversions presented in our analysis compared to those derived from the standard HMI ring-analysis pipeline inversion.
In our analysis, 
the logarithmic gradient is close to zero around 1.6 Mm and becomes steeper towards the surface --- layer S.
However, around a depth of 1 Mm, there is a lack of consistency between OLA and RLS results; there are also differences between the northern and southern hemispheres, as indicated in Figure~\ref{fig:sym_deep} (note the large north-south antisymmetric component  at this depth). To us this
indicates that neither 15-degree nor 30-degree tiles, give reliable results at very shallow depths, even after averaging the data over 12 years.
This implies that layer S is poorly characterized when using 15 and 30-degree tiles.

 %==========================================================
 \begin{table}
\begin{center}
 \caption{
  Parameters of the minimum in the gradient of the rotation: its location, amplitude, and full width at half maximum (FWHM).
}
 \label{tab:minimum}
 \begin{tabular}{l||rr|rr}
 Parameter & OLA  & RLS   & OLA & RLS \\
its uncertainty & 30-deg & tiles  & 15-deg & tiles\\
\hline
\hline
Depth [Mm] & 2.44 & 2.44 & 2.7 & 2.7 \\
 & 0.07 & 0.07 & 0.1 & 0.1 \\
 \hline
Amplitude & $-2.7$ & $-3.2$ & $-2.6$ & $-2.6$ \\
($\partial \ln \Omega/\partial \ln r$) & 0.1 & 0.4 & 0.2 & 0.2 \\
 \hline
FWHM [Mm] & 1.0 & 1.0 & 1.3 & 1.7 \\
 & 0.1 & 0.1 & 0.1 & 0.2 \\
 \hline
 \end{tabular}
 \end{center}
\end{table}
%==========================================================

\subsection{The latitudinal variation of the NSSL}

As mentioned in Section~1, earlier investigations find only a small variation of the gradient with latitude, while others report 
the gradient to be zero or even positive at high latitudes. The ring-diagram analysis method allows for a more detailed examination of the rotational gradient near the solar surface compared to global helioseismology.
However, its uncertainties exceed those reported in global helioseismology studies, posing challenges
in detecting potential variations with latitude.

\begin{figure}
	\includegraphics[width=\columnwidth]{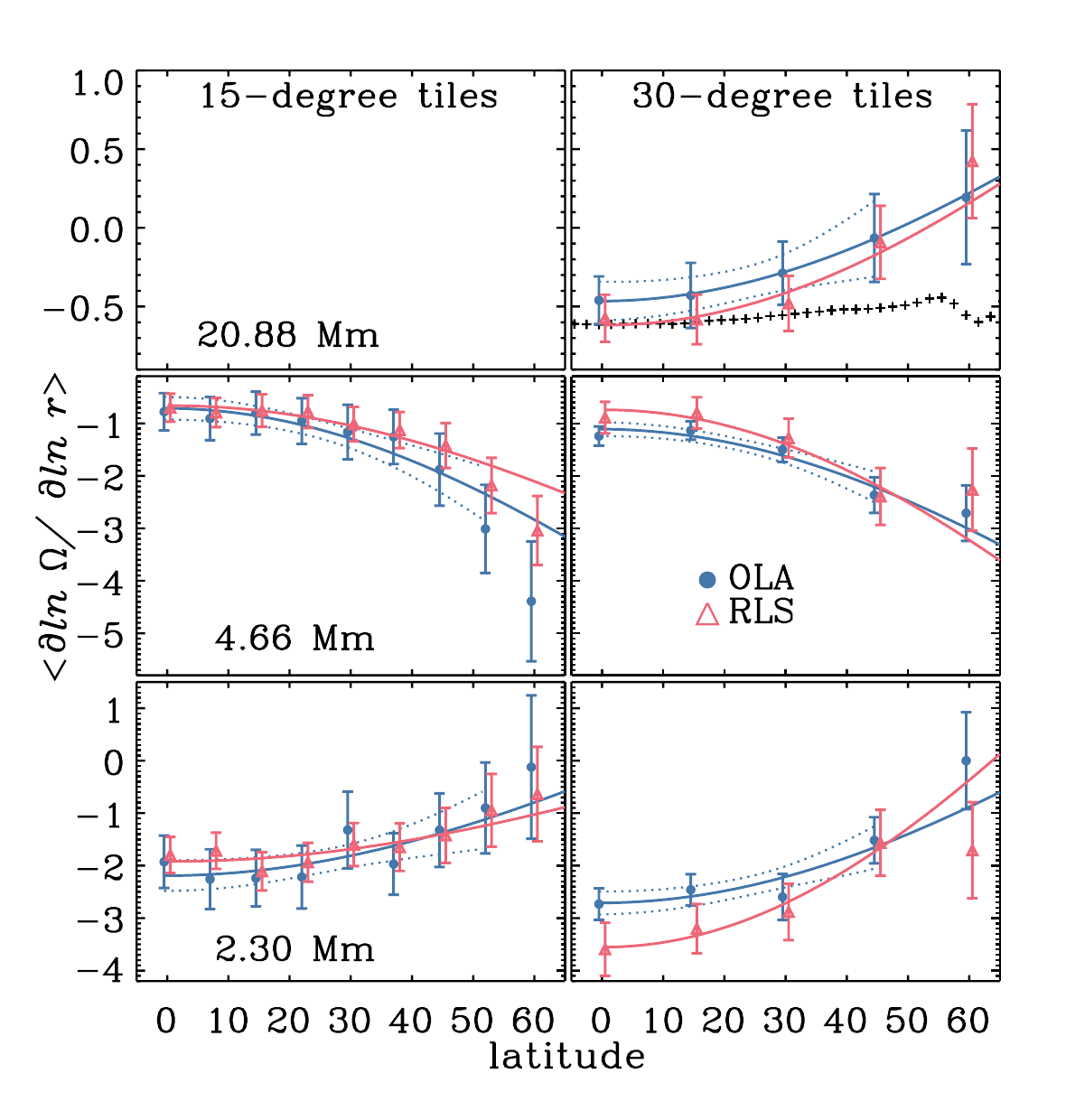}
    \caption{
    The North-South symmetric component of the rotational gradient plotted against latitude at three depths: 20.88 Mm (0.970 R$_\odot$), 4.66 Mm (0.9933 R$_\odot$), and 2.30 Mm (0.9967 R$_\odot$) for both inversion methods and tile sizes.
    The solid lines represent the linear fits against cosine(latitude) for latitudes smaller than 60$^\circ$. While values for 60$^\circ$ latitude are shown, they were not included in the fitting. Dotted lines indicate the confidence level of the fitting to the OLA results. Those for the RLS fitting have similar values.
    The black crosses in the top right panel represent results obtained by \citet{antia2022} for HMI.
    }
    \label{fig:gradient985_latitude}
\end{figure}

To investigate the latitude-dependent variation of the rotational gradient in our results, we fitted a straight line to the north-south symmetric gradient as a function of the cosine of latitude.
We use results for latitudes below $60^\circ$, due to substantial oscillations in our $60^\circ$ solution, as can be seen in Figure~\ref{fig:results15_30_96}.
For both OLA and RLS, we could identify three depth ranges that show Pearson correlation coefficients that indicate a probability of lower than 0.05 for no correlation: 19.5 $\pm$ 2.1 Mm, 4.11 $\pm$ 0.70 Mm, and 2.23 $\pm$ 0.14 Mm. Additionally, the slope exceeds 1.4$\sigma$, signifying a probability of 85\% or more that the correlation coefficient is non-zero.
The Pearson correlation coefficients with absolute values exceeding 0.95 for the 30-degree tiles (four latitudes) and 0.707 for the 15-degree tiles (eight latitudes) are statistically  significant.
Figure~\ref{fig:gradient985_latitude} illustrates the latitudinal variation in these three depth ranges for both the OLA and RLS inversion methods and different tile sizes.

Note that for two of the three depth ranges, the gradient is zero or even positive at high latitudes. In the top panel, at a radius of 0.97 R$_\odot$ (depth of 20.88 Mm), crosses represent results obtained by \citet{antia2022} for HMI. 
While our findings deviate at high latitudes, they remain within 1.5$\sigma$ of the uncertainty. \citet{barekat2014,barekat2016} also show only a slight increase of the gradient with latitude, which becomes less negative, akin to \citet{antia2022}.
In contrast to these earlier results, our results exhibit a change towards positive values with increasing latitude, similar to \citet{corbard2002} and \citet{zaatri2009}, although their measurements of the variation with latitude are at a different radius, around 0.99 R$_\odot$.
At 0.99 R$_\odot$, we obtained a substantial correlation coefficient of $-0.94$ for the 15-degree tiles, indicating a probability of less than 0.001 (0.1\%) of no correlation. However, the associated uncertainty brings the slope within the 1$\sigma$ range, indicating that it is statistically consistent with zero.

We also observe a null gradient at high latitudes at a very shallow depth (bottom panel of Fig.~\ref{fig:gradient985_latitude}). Our results agree with \citet{reiter2020} at 30$^\circ$ latitude and higher, obtained very close to the solar surface (see their Fig. 28).
Nonetheless, as depicted in Fig.~\ref{fig:results15_30_96}, this phenomenon is
associated with the widening and inward displacement of the Middle shear layer at latitudes $\ge$ 45$^\circ$.
At a depth closer to the dip in the gradient, around $4$ Mm, we observe a positive linear correlation with cosine of latitude (i.e., a negative correlation with latitude), where the gradient is even steeper at high latitude (shown in Fig.~\ref{fig:gradient985_latitude} middle panel). 
However, this result is also associated with the inward displacement of the Middle shear layer at latitudes $\ge$ 45$^\circ$ seen in Fig.~\ref{fig:results15_30_96}.
This unexpected change underscores the intricacies of solar dynamics at this specific depth and warrants further analysis.

%%%%%%%%%%%%%%%%%%%%%%
%%%%%%%%%%%%%%%%%%%%%%

 \subsection{Further analysis of Layer M}

\begin{figure}
	\includegraphics[width=\columnwidth]{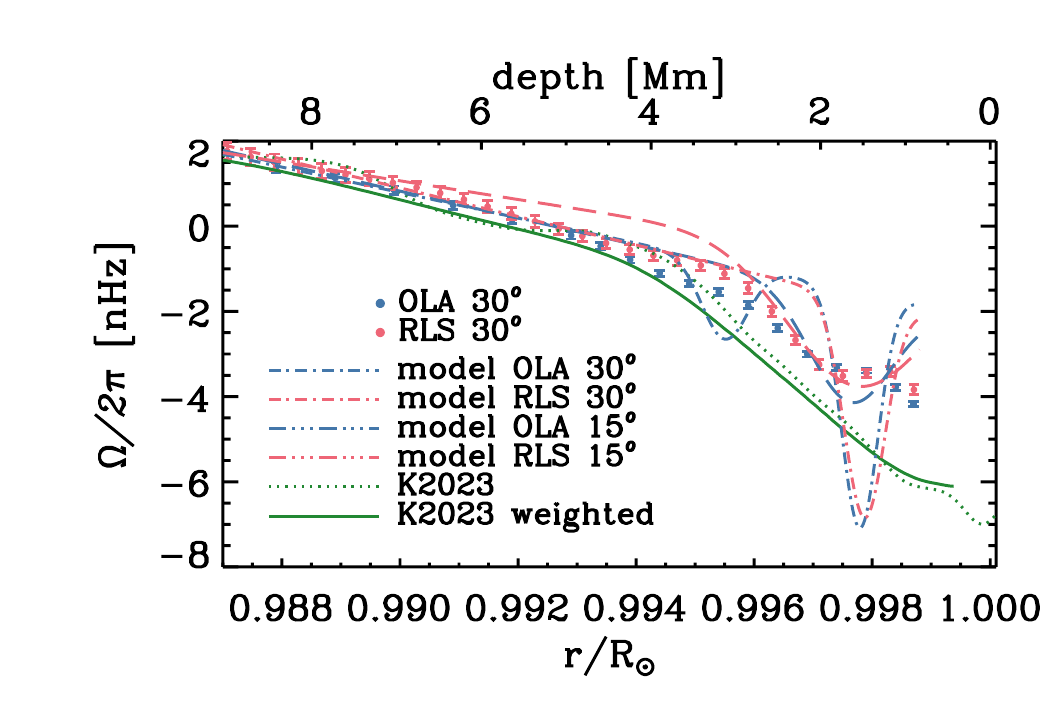}
    \includegraphics[width=\columnwidth]{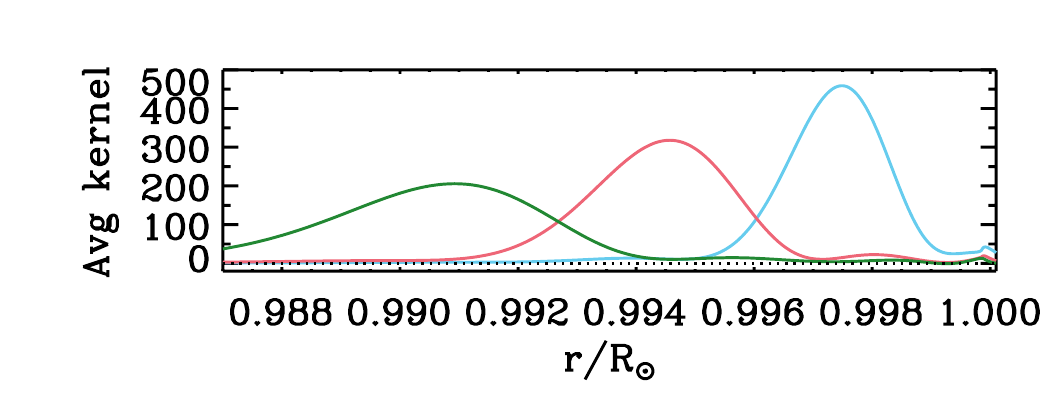}
    \caption{Top panel. A comparison of solar rotation rates, obtained by inversion results for the 30-degree tiles (full circles), a simple model for the 15 and 30-degree tiles and simulations, is shown after subtracting the 
    Carrington rotation rate. The rotation rate estimated by \citet[]{irina2023} using RHD simulations is represented by the dotted green line. The full green line shows their results weighted by the OLA averaging kernels for the 30-degree tiles at the equator. Both lines are arbitrarily displaced by 2.9 nHz to align with our results for r $<$ 0.992 R$_\odot$. Bottom panel. Three examples of OLA averaging kernels in different colors for the 30-degree tiles at the equator are shown to illustrate the challenge of reproducing the model in the inversion results. The horizontal dotted line indicates a zero value for the averaging kernel.
    }
    \label{fig:uxmodel}
\end{figure}

Given the large width of the averaging kernels, the observed rapid increase in the shear within layer M, likely occurs much closer to the solar surface than is indicated by the inversion results. Based on our inversion results, we assume a model for the zonal flow that incorporates a Gaussian
with negative amplitude centered around 0.996 R$_\odot$ (2.8 Mm), along with a straight line. The model involves five parameters --- amplitude, width, and location for the Gaussian, a slope, and a constant. Before adjusting the parameters using a Levenberg-Marquardt least-squares method to match the inverted values of $\Omega$ obtained by OLA and RLS at the equator, the model was weighted by the averaging kernels obtained by the OLA method. For the OLA 30-degree case, a second Gaussian at around 0.994$R_\odot$ was included in the model.
The center of the Gaussian could be determined very accurately, and the fit residuals (differences in $\Omega/2\pi$ between observations and the model)
were smaller than 0.1 nHz.
The residuals exhibited an oscillatory pattern, characteristic of the error correlation inherent in the inversion results. We estimated the gradient from the fitted $\Omega$, which showed agreement with observed values, with differences smaller than 0.4 --- below observed uncertainties, and hence statistically insignificant. 
The relevant fitted parameters are presented in Table~\ref{tab:fittedux}.
A visual representation of the models obtained from fitting two tile sizes and two inversion methods are shown in Fig.~\ref{fig:uxmodel}. In this figure, the results are plotted at the center of gravity of the averaging kernel, explaining why the rotation rates do not go closer to the surface.

In conclusion, our estimates of the near-surface dimensionless gradient of rotation  is consistent with a thin layer approximately 1 Mm wide (0.6 Mm for the 30-degree and 1.5 Mm for the 15-degree case) where there is a sudden decrease in the rotation rate as one goes outwards, decreasing by 3-5 nHz (5 nHz for the 30-degree and 3 nHz for the 15-degree case). This layer is located at a depth of approximately 1.5 Mm. 
Although the minimum location in the observed rotation gradient differs slightly between the 30-degree and 15-degree tiles, the locations of the fitted Gaussians remain consistent. In Fig.~\ref{fig:uxmodel}, we observe that due to the large width of the averaging kernels, the effect of the sharp decrease in rotation rate is seen at much deeper layers, starting around 3 Mm deep, while the minimum is, in fact, at 1.5 Mm. 
The maximum gradient, in terms of its absolute value, corresponds to the middle of the decline in the Gaussian. After the minimum, rotation appears to increase again, but results shallower than $\sim$ 1 Mm are not considered trustworthy.

When applying OLA averaging kernels obtained with the 30-degree tiles for latitudes other than the Equator to the Gaussian model shown in Fig.~\ref{fig:uxmodel}, the rotation gradient remains essentially the same, with the exception that the amplitude of the minimum decreases with increase in latitude. It reaches approximately 80\% of the value at the Equator for latitudes 45$^\circ$ and 60$^\circ$. A similar behavior is observed when applying OLA 15-degree averaging kernels --- the amplitude of the minimum decreases by as much as 15\% at high latitudes. 
We find larger variations in the observed rotational gradient at high latitudes ($>$ 45$^\circ$) than at lower latitudes, and no well-defined minimum or displacement towards the interior. This suggests that our data may not be good enough at these latitudes due to foreshortening.

 \begin{table}
\begin{center}
 \caption{
Fitted model parameters for $\Omega$: Gaussian center, height, and full width at half maximum (FWHM).
}
 \label{tab:fittedux}
 \begin{tabular}{l||rr|rr}
Parameter & OLA & RLS & OLA & RLS \\
its uncertainty &  30-deg &  tiles &  15-deg &  tiles \\
\hline
\hline
Depth [Mm] &  1.5  &  1.48   &  1.65  &  1.6  \\
 &  0.3 & 0.02 &  0.05 &  0.1 \\
 \hline
Height [nHz] &  $-6$  &  $-5$ &  $-2.6$ &  $-3.1$ \\
 &  9 &  1 &  0.3 &  0.2\\
 \hline
FWHM [Mm] &  0.5  &  0.6   &  1.1   &  1.9   \\
 &  1.0 &   0.2 &  0.3 &   0.3 \\
 \hline
 \end{tabular}
 \end{center}
\end{table}

%%%%%%%%%%%%%%%%%%%%%%
%%%%%%%%%%%%%%%%%%%%%%

\begin{figure}
	\includegraphics[width=\columnwidth]{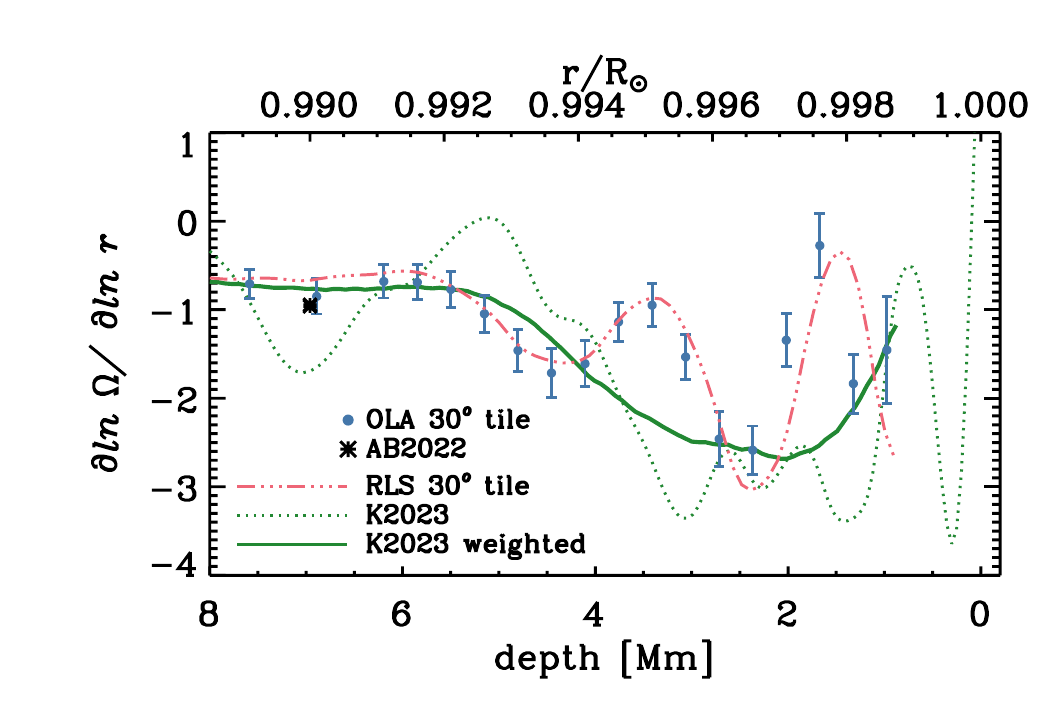}
    \caption{
    The dimensionless gradient of the rotation rate obtained by \citet[]{irina2023} for a latitude of 30$^\circ$, depicted by the dotted green line. The full green line represents the gradient after applying OLA averaging kernels. To facilitate comparison with our results, the Carrington rotation rate was added to the simulation rotation rate before calculating the logarithm.
    The gradient derived from OLA results are shown as full blue circles, while RLS results are shown as a triple-dot-dashed red line; both results are for latitude 30$^\circ$ North. The results for 30$^\circ$ South are not displayed since they closely mirror those for 30$^\circ$ North. The uncertainties of OLA and RLS results are comparable, and hence, error bars are only shown for OLA.
    Global helioseismology results \citep[]{antia2022} at the same latitude and at 0.99 R$_\odot$ is represented as a black star. The error bars for these results are too small to be visible, and the stars correspond to the overlapped results for GONG, MDI, and HMI for cycles 23 and 24 when available.
    }
    \label{fig:irina_dlnomega}
\end{figure}

\citet[]{irina2023} conducted a detailed analysis of the impact of rotation within the solar convection zone, employing 3D radiative hydrodynamic (RHD) modeling. The study focused on the upper layers of solar convection within a local domain measuring 80 Mm in width (80 × 80 Mm) and 25 Mm in depth, situated at 30$^\circ$ latitude.
By averaging the azimuthal flows over both horizontal directions and a 24-hour timeframe \citep[depicted in Fig. 3a of][and reproduced in Fig.~\ref{fig:uxmodel} by the dotted green line]{irina2023}, the study observed notable deviations from the imposed rotation rate. 
To compare their findings with ours, we applied averaging kernels obtained from OLA inversions for the 30-degree tiles at the Equator --- the differences between the averaging kernels at the Equator and 30$^\circ$ latitudes are very small. As expected, the large width of these averaging kernels smoothed out the simulation results.
Fig.~\ref{fig:irina_dlnomega} displays the gradient of the rotation rate obtained by \citet[]{irina2023} before and after applying our OLA averaging kernels. As expected, the averaging kernels smooth out their gradient, but it reproduces well the layer between depths of 4 Mm and 1 Mm with the largest gradient, in absolute value --- closely coinciding with the depth limits of layer M. Generally, their findings (weighted by the same averaging kernels as ours) align with ours.
Crucially, the minimum of the gradient is located at the same layer as in our results, approximately 3 Mm deep. However, our results revealed a much narrower layer with a markedly negative gradient, less than half of the width of what the simulation indicated.

%%%%%%%%%%%%%%%%%%%%%%
%%%%%%%%%%%%%%%%%%%%%%

\subsection{The NSSL away from the Central Meridian}

Helioseismic measurements, including fitted flows from ring-diagram analysis and travel time differences from time-distance analysis, made at heliocentric longitudes away from the central meridian are known to be affected by systematic discrepancies that cannot be physical
\citep[][among others]{braun2008,duvall2009,zhao2012}.
\citet{zhao2012} associated a systematic East-West trend in the zonal flows along the equator seen in time-distance results with an inferred center-to-limb effect.
They used this to correct the North–South measurements obtained along the central meridian, leading to a more consistent determination of meridional flows within the solar interior and thereby justifying their correction.
\citet{baldner2012}, employing numerical simulations of convection \citep{stein2009}, demonstrated that vertical flows from convection could indeed explain such a center-to-limb effect. 
They concluded that the ad hoc correction implemented by \citet{zhao2012}, now widely used, is probably warranted. 
\citet{woodard2013} highlighted that a radial variation in the phase of the mode eigenfunctions, in combination with a center-to-limb variation in the height at which the oscillations are observed, would also introduce errors in the measured flow.

\begin{figure*}
    \includegraphics[width=\textwidth]{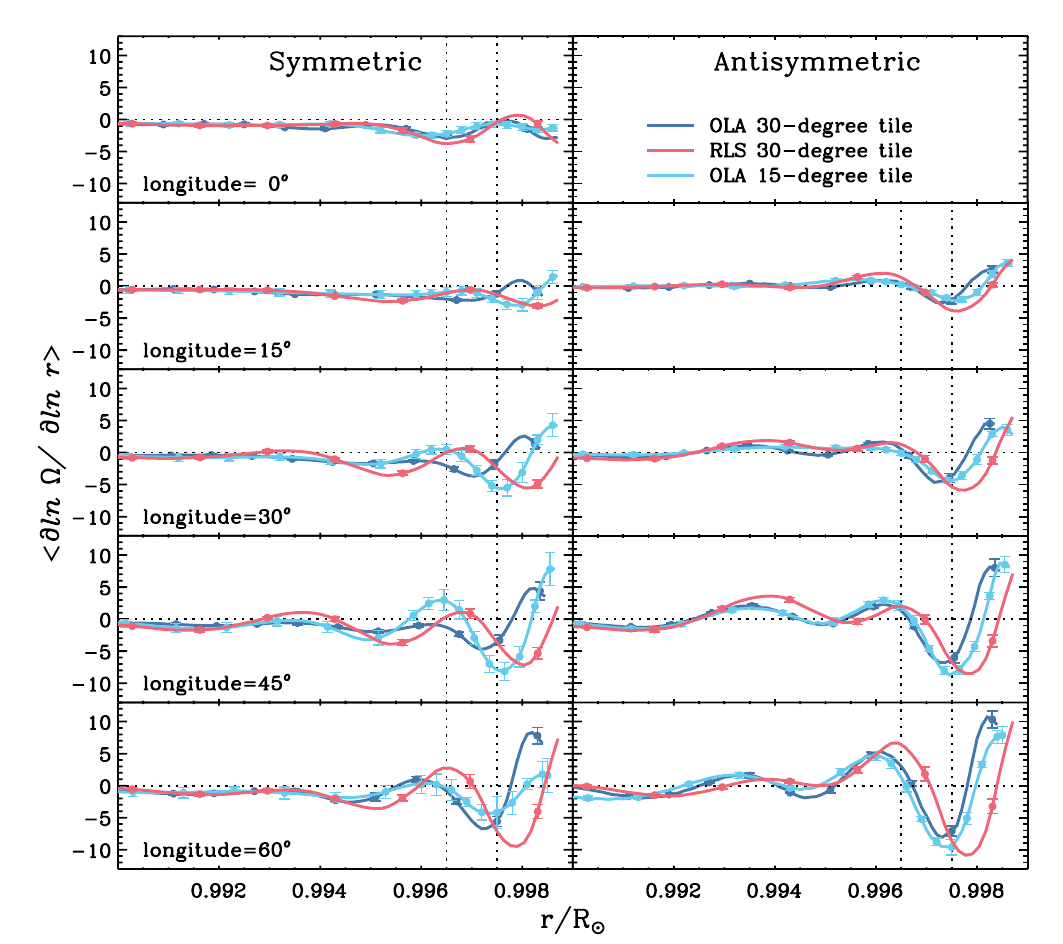}
    \caption{  
    East-West symmetric (left) and antisymmetric (right) components of the rotational gradient, obtained for patches centered at the equator and various Stonyhurst longitudes. Results using RLS inversion on 15-degree tiles were omitted for better visualization. 
    The uncertainty is shown for a few radii. 
    The vertical line at 0.9965 R$_\odot$ (a depth of $2.4$ Mm) corresponds to the minimum gradient observed at the central meridian.
    }
    \label{fig:alongEQ}
\end{figure*}

In this paper, we have so far focused our analysis exclusively on regions at the central meridian. The results of previous studies of \citet{zaatri2009,komm2022} were shown after averaging over Stonyhurst longitude, and some \citep[e.g.,][]{komm2022} applied a correction for systematic variations with disk position before averaging.
To assess possible systematic effects that might affect our measurements at different latitudes, we also examined results along the equator.
Figure~\ref{fig:alongEQ} illustrates the symmetric and antisymmetric components of the rotation gradient along the solar equator at various longitudes relative to central meridian. The symmetric component shows a distinct minimum around a depth of 3 Mm at the central meridian. Away from it, larger variations are observed, but there is no agreement among different tile sizes and inversion methods. This suggests that any center-to-limb effect is sensitive to both tile size and inversion technique. In contrast, the variations of the antisymmetric component of the gradient ((W-E)/2) remain consistent across inversion methods and tile sizes. The amplitude of these variations increases with longitude, with a maximum at 1.7 Mm (0.9975 R$_\odot$) for all longitudes. The presence of a systematic anti-symmetric component cannot be attributed to a center-to-limb effect.
The North minus South logarithmic gradient differences do not exhibit a steady pattern with radius, as observed for the W-E differences, and are much smaller, comprising less than 10\% of the values observed in the right panel.

\begin{figure}
 \includegraphics[width=\columnwidth]{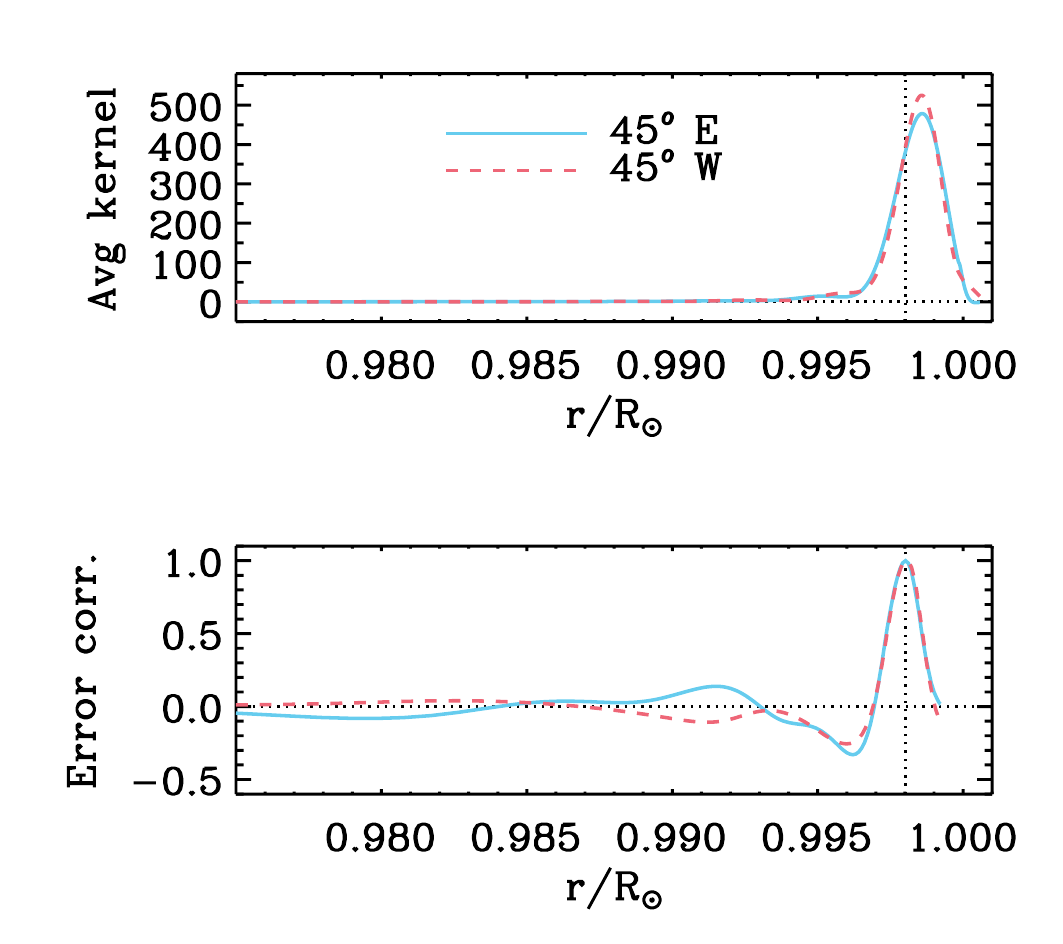}
    \caption{
    Averaging kernels (top) and error correlations (bottom)  for the 30-degree tiles at 0.998 R$_\odot$ at the Equator and Stonyhurst longitude 45$^\circ$E (blue) and 45$^\circ$W (red). As usual, the averaging kernels are normalized such that their integral over radius is 1.
    }
    \label{fig:avk_alongEQ}
\end{figure}

Figure~\ref{fig:avk_alongEQ} presents a comparison of the averaging kernels and error correlations between 45$^\circ$ Stonyhurst longitudes at East and West,
at a depth where substantial differences exist in the rotation gradient,
revealing minimal differences. They also agree well with those at central meridian and different latitudes (shown in Fig. 1).
Therefore, neither can explain the East-West differences.
The east-west antisymmetry induced by solar rotation likely contributes to this phenomenon; however, its exact origin remains unknown.

%%%%%%%%%%%%%%%%%%%%%%
%%%%%%%%%%%%%%%%%%%%%%

%%%%%%%%%%%%%%%%%%%%%%

\section{Discussion and Conclusions}

We have determined the dimensionless gradient, $\partial \ln \Omega/\partial \ln r$, of the rotation rate in the near-surface shear layer of the Sun using ring-diagram analyses applied to twelve years of HMI data. We have not relied on the HMI-pipeline inversions; instead we have done a thorough and meticulous analysis that has resulted in estimates that are more precise and accurate than earlier results obtained with local helioseismic analyses.
We took into account
the variation of the logarithmic gradient with radius and the anticipated oscillations in the inversion solution due to error correlation when examining the gradient's latitude variation at a given radius.

Our results show that the NSSL, generally thought to constitute a simple shear layer, actually has substructure (Fig.~\ref{fig:layers}). In the deeper layers (layer D), 
the dimensionless gradient within the NSSL depends on radius.
The weighted average of the gradient obtained by OLA and RLS and 
for latitudes $\leq~30^\circ$ steepens in value from $-0.49\pm$0.07 at a depth of 21 Mm to $-0.81\pm$0.07 at 11 Mm. It is then almost constant until a depth of 6 Mm (0.992 R$_\odot$), with an average value of $-0.7\pm$0.2 (including both OLA and RLS results). Closer to the surface, it steepens further, to  approximately $-1$ at a depth of 4 Mm (0.995 R$_\odot$), followed by a considerably faster change, culminating in the largest negative gradient at a depth 3 Mm with an amplitude approximately three times larger;
this we refer to as layer M.
Even closer to the surface, depths smaller than 3 Mm, a significant change in the gradient 
is observed, and the logarithmic gradient approaches zero at approximately 1.6 Mm. We call depths smaller than $\sim$1.6 Mm layer S.
Our results are unreliable at layers shallower than $\sim$1 Mm.

In layer M, the solar rotation rate decreases rapidly with decreasing depth from around 3 Mm depth, dropping by 2 nHz (or more) at about 2 Mm deep. It then stabilizes toward the surface, at least until around 1 Mm deep. 
To assess the impact of the width of the averaging kernels  on our estimates of the rotation rate, we used a simple model and examined how it would be reproduced in our analysis. The  model consisted of a linear component and a Gaussian with a negative amplitude. This analysis suggests that the solar rotation rate declines towards the surface more sharply than the inversion results, starting at a shallower depth, around 2 Mm deep, with a decrease of 5 nHz at 1.5 Mm depth. After reaching this minimum, the rotation rate appears to increase again; note that  at this point, we are either within, or in close proximity to, the super-adiabatic layer.
We also applied the averaging kernels to the rotation rate calculated from \citet{irina2023}'s 3D radiative simulation. Their simulation too exhibits a second shear layer (layer M) whose rotational gradient has similar amplitude and location to that observed in this work, but with a much broader width.

Additionally, while our analysis concentrated  on regions at the central meridian, we also investigated systematic errors at longitudes off the central meridian.
The antisymmetric component of the gradient, (W-E)/2, exhibits a significant difference around a depth of 1.7 Mm, with its amplitude increasing as longitude rises. This behavior differs markedly from the N-S antisymmetric component of the gradient, suggesting that the observed latitude variations of the rotation gradient are not strongly affected by a center-to-limb effect.
It is possible that effects of tracking solar rotation in the ring-diagram analysis introduce the asymmetry in the E-W direction, for which there is no ready physical explanation. We must be cautious about ruling out potential systematic effects in determinations of variations of the rotational gradient with latitude.
Therefore, we have refrained from making any ad hoc corrections that might introduce a spurious signal in our latitude results.

Our findings corroborate earlier observations from various studies
\citep{basu_etal1999, zaatri2009, komm2022}, confirming the presence of a
second shear layer located closer to the solar surface (layer M). 
This layer exhibits a significantly stronger shear rate compared to the one typically identified as the NSSL (referred to as layer~D here).
\citet{komm2022} in a recent work reported a change in the steepness of the radial gradient between 9.7 and 1.4 Mm deep, where it changes from $\approx -1$ to $-2.6$. The largest amplitude agrees well with ours, but its reported location is much shallower.
In this study, we achieved a more precise determination of the location of the gradient steepening compared to previous works.

There have been contradictory results about  the variation of the rotational gradient with solar latitude.
While some studies, such as those by \citet{barekat2014, barekat2016} and \citet{antia2022}, observed only a minor decrease in the amplitude of the gradient with solar latitude, others \citep{corbard2002,zaatri2009} reported a significant decrease.
We observe a correlation between the logarithmic gradient and latitude for both inversion methods and tile sizes, albeit confined to specific radius ranges. However, it is essential to acknowledge the possibility of small latitude variations elsewhere given the uncertainties inherent in gradient determination. Within the depth range of 21.6 to 17.4 Mm (0.969-0.975 R$_\odot$), a substantial increase in the logarithmic gradient with latitude is noted, aligning with the findings of \citet{corbard2002} and \citet{zaatri2009} when analyzing radii closer to the surface than ours ($\sim$ 7 Mm deep).

The NSSL, and in particular layers M and S, lies in an interesting part of the Sun in terms of structure. If we consider a standard solar model, such as 
Model S \citep[Model S ---][]{jcd1996}, significant changes in structure are seen below approximately 7 Mm, coinciding with the occurrence of layers M and S. 
The density 
changes by two orders of magnitude between depths of 6 Mm to 1 Mm,
transitioning from $2 \times 10^{-4}$ g/cm$^{3}$ at 6 Mm to $2 \times 10^{-6}$ g/cm$^{3}$ around 1 Mm deep, based on Model S.
This region marks the change from the HeII ionization zone to the HeI and HI ionization zones, accompanied by the related sharp decrease in the adiabatic index $\Gamma_1$ towards the surface. The width of the layer of slow rotation (layer M) is comparable to the pressure and density scale heights, which vary within (0.8$\pm$0.3) Mm and (1.0$\pm$0.4) Mm, respectively. We observed another change in the rotational gradient (layer S) even closer to the surface and, 
in the outermost 1 Mm, as helium and hydrogen become neutral, there is another abrupt change in $\Gamma_1$, with a sharp increase towards the solar surface.
Also, a layer characterized by considerable superadiabaticity emerges. As radiative energy transport becomes dominant, the efficiency of convective energy flux transport diminishes.

Due to the rapid decrease in density, the angular momentum diminishes to a quarter of its value between depths of 13 Mm and 7 Mm, and to less than 1\% from 7 Mm to 1 Mm. The observed variation of $\Omega$(r) with radius has a limited impact on angular momentum, given that the variation of moment of inertia with depth exhibits a much steeper trend compared to the rotational variation.
In most of the NSSL, specifically in layer D, the decrease in the rotation rate towards the surface corresponds to a decrease of $\sim$0.1\% per Mm in the angular momentum in relation to a constant rotation rate, while in layer M, which has a stronger shear, the angular momentum decreases by $\sim$0.3\% per Mm.

Our results, therefore, show that there is still much to learn about the NSSL. In particular, the co-location of different sub-layers of the NSSL with significant structural features suggests a deeper relation between the two, and needs to be studied further.

\begin{acknowledgments}
This research was supported in part by NASA Contract NAS5-02139 to Stanford University.
This work uses data from the Helioseismic and Magnetic Imager. HMI data are courtesy of NASA/SDO and the HMI science team. The data used in this article are publicly available from the Joint Science Operations Center at jsoc.stanford.edu.
\end{acknowledgments}

\vspace{5mm}
\facilities{HMI(SDO), JSOC(Stanford and Lockheed)}

\bibliography{main}{}

\begin{thebibliography}{}
\expandafter\ifx\csname natexlab\endcsname\relax\def\natexlab#1{#1}\fi
\providecommand{\url}[1]{\href{#1}{#1}}
\providecommand{\dodoi}[1]{doi:~\href{http://doi.org/#1}{\nolinkurl{#1}}}
\providecommand{\doeprint}[1]{\href{http://ascl.net/#1}{\nolinkurl{http://ascl.net/#1}}}
\providecommand{\doarXiv}[1]{\href{https://arxiv.org/abs/#1}{\nolinkurl{https://arxiv.org/abs/#1}}}

\bibitem[{{Antia} \& {Basu}(2022)}]{antia2022}
{Antia}, H.~M., \& {Basu}, S. 2022, \apj, 924, 19, \dodoi{10.3847/1538-4357/ac32c3}

\bibitem[{{Backus} \& {Gilbert}(1968)}]{backus1968}
{Backus}, G., \& {Gilbert}, F. 1968, Geophysical Journal, 16, 169, \dodoi{10.1111/j.1365-246X.1968.tb00216.x}

\bibitem[{{Baldner} \& {Schou}(2012)}]{baldner2012}
{Baldner}, C.~S., \& {Schou}, J. 2012, \apjl, 760, L1, \dodoi{10.1088/2041-8205/760/1/L1}

\bibitem[{{Barekat} {et~al.}(2014){Barekat}, {Schou}, \& {Gizon}}]{barekat2014}
{Barekat}, A., {Schou}, J., \& {Gizon}, L. 2014, \aap, 570, L12, \dodoi{10.1051/0004-6361/201424839}

\bibitem[{{Barekat} {et~al.}(2016){Barekat}, {Schou}, \& {Gizon}}]{barekat2016}
---. 2016, \aap, 595, A8, \dodoi{10.1051/0004-6361/201628673}

\bibitem[{{Basu} \& {Antia}(1999)}]{basu_antia1999}
{Basu}, S., \& {Antia}, H.~M. 1999, \apj, 525, 517, \dodoi{10.1086/307900}

\bibitem[{{Basu} {et~al.}(1999){Basu}, {Antia}, \& {Tripathy}}]{basu_etal1999}
{Basu}, S., {Antia}, H.~M., \& {Tripathy}, S.~C. 1999, \apj, 512, 458, \dodoi{10.1086/306765}

\bibitem[{{Bogart} {et~al.}(2011){Bogart}, {Baldner}, {Basu}, {Haber}, \& {Rabello-Soares}}]{Bogart_pipeline}
{Bogart}, R.~S., {Baldner}, C., {Basu}, S., {Haber}, D.~A., \& {Rabello-Soares}, M.~C. 2011, in Journal of Physics Conference Series, Vol. 271, GONG-SoHO 24: A New Era of Seismology of the Sun and Solar-Like Stars, 012008, \dodoi{10.1088/1742-6596/271/1/012008}

\bibitem[{{Braun} \& {Birch}(2008)}]{braun2008}
{Braun}, D.~C., \& {Birch}, A.~C. 2008, \apjl, 689, L161, \dodoi{10.1086/595884}

\bibitem[{{Christensen-Dalsgaard} {et~al.}(1996){Christensen-Dalsgaard}, {Dappen}, {Ajukov}, {Anderson}, {Antia}, {Basu}, {Baturin}, {Berthomieu}, {Chaboyer}, {Chitre}, {Cox}, {Demarque}, {Donatowicz}, {Dziembowski}, {Gabriel}, {Gough}, {Guenther}, {Guzik}, {Harvey}, {Hill}, {Houdek}, {Iglesias}, {Kosovichev}, {Leibacher}, {Morel}, {Proffitt}, {Provost}, {Reiter}, {Rhodes}, {Rogers}, {Roxburgh}, {Thompson}, \& {Ulrich}}]{jcd1996}
{Christensen-Dalsgaard}, J., {Dappen}, W., {Ajukov}, S.~V., {et~al.} 1996, Science, 272, 1286, \dodoi{10.1126/science.272.5266.1286}

\bibitem[{{Corbard} \& {Thompson}(2002)}]{corbard2002}
{Corbard}, T., \& {Thompson}, M.~J. 2002, \solphys, 205, 211, \dodoi{10.1023/A:1014224523374}

\bibitem[{{Duvall} \& {Hanasoge}(2009)}]{duvall2009}
{Duvall}, T.~L., J., \& {Hanasoge}, S.~M. 2009, in Astronomical Society of the Pacific Conference Series, Vol. 416, Solar-Stellar Dynamos as Revealed by Helio- and Asteroseismology: GONG 2008/SOHO 21, ed. M.~{Dikpati}, T.~{Arentoft}, I.~{Gonz{\'a}lez Hern{\'a}ndez}, C.~{Lindsey}, \& F.~{Hill}, 103, \dodoi{10.48550/arXiv.0905.3132}

\bibitem[{{Hill}(1988)}]{hill1988}
{Hill}, F. 1988, \apj, 333, 996, \dodoi{10.1086/166807}

\bibitem[{{Howe} \& {Thompson}(1996)}]{howe_thompson}
{Howe}, R., \& {Thompson}, M.~J. 1996, \mnras, 281, 1385, \dodoi{10.1093/mnras/281.4.1385}

\bibitem[{{Karak} \& {Cameron}(2016)}]{karak2016}
{Karak}, B.~B., \& {Cameron}, R. 2016, \apj, 832, 94, \dodoi{10.3847/0004-637X/832/1/94}

\bibitem[{{Kitiashvili} {et~al.}(2023){Kitiashvili}, {Kosovichev}, {Wray}, {Sadykov}, \& {Guerrero}}]{irina2023}
{Kitiashvili}, I.~N., {Kosovichev}, A.~G., {Wray}, A.~A., {Sadykov}, V.~M., \& {Guerrero}, G. 2023, \mnras, 518, 504, \dodoi{10.1093/mnras/stac2946}

\bibitem[{{Komm}(2022)}]{komm2022}
{Komm}, R. 2022, Frontiers in Astronomy and Space Sciences, 9, 428, \dodoi{10.3389/fspas.2022.1017414}

\bibitem[{{Parker}(1975)}]{parker1975}
{Parker}, E.~N. 1975, \apj, 198, 205, \dodoi{10.1086/153593}

\bibitem[{{Reiter} {et~al.}(2020){Reiter}, {Rhodes}, {Kosovichev}, {Scherrer}, {Larson}, \& {}}]{reiter2020}
{Reiter}, J., {Rhodes}, E.~J., J., {Kosovichev}, A.~G., {et~al.} 2020, \apj, 894, 80, \dodoi{10.3847/1538-4357/ab7a17}

\bibitem[{{Rhodes} {et~al.}(1990){Rhodes}, {Cacciani}, \& {Korzennik}}]{rhodes1990}
{Rhodes}, E.~J., {Cacciani}, A., \& {Korzennik}, S.~G. 1990, in Progress of Seismology of the Sun and Stars, ed. Y.~{Osaki} \& H.~{Shibahashi}, Vol. 367 (Springer-Verlag), 163, \dodoi{10.1007/3-540-53091-6_77}

\bibitem[{{Scherrer} {et~al.}(2012){Scherrer}, {Schou}, {Bush}, {Kosovichev}, {Bogart}, {Hoeksema}, {Liu}, {Duvall}, {Zhao}, {Title}, {Schrijver}, {Tarbell}, \& {Tomczyk}}]{scherrer2012}
{Scherrer}, P.~H., {Schou}, J., {Bush}, R.~I., {et~al.} 2012, \solphys, 275, 207, \dodoi{10.1007/s11207-011-9834-2}

\bibitem[{{Sekii}(1997)}]{sekii}
{Sekii}, T. 1997, in Sounding Solar and Stellar Interiors, ed. J.~{Provost} \& F.-X. {Schmider}, Vol. 181, ISBN0792348389

\bibitem[{{Spiegel} \& {Zahn}(1992)}]{spiegel1992}
{Spiegel}, E.~A., \& {Zahn}, J.~P. 1992, \aap, 265, 106

\bibitem[{{Stein} {et~al.}(2009){Stein}, {Nordlund}, {Georgoviani}, {Benson}, \& {Schaffenberger}}]{stein2009}
{Stein}, R.~F., {Nordlund}, {\r{A}}., {Georgoviani}, D., {Benson}, D., \& {Schaffenberger}, W. 2009, in Astronomical Society of the Pacific Conference Series, Vol. 416, Solar-Stellar Dynamos as Revealed by Helio- and Asteroseismology: GONG 2008/SOHO 21, ed. M.~{Dikpati}, T.~{Arentoft}, I.~{Gonz{\'a}lez Hern{\'a}ndez}, C.~{Lindsey}, \& F.~{Hill}, 421

\bibitem[{{Woodard} {et~al.}(2013){Woodard}, {Schou}, {Birch}, \& {Larson}}]{woodard2013}
{Woodard}, M., {Schou}, J., {Birch}, A.~C., \& {Larson}, T.~P. 2013, \solphys, 287, 129, \dodoi{10.1007/s11207-012-0075-9}

\bibitem[{{Zaatri} \& {Corbard}(2009)}]{zaatri2009}
{Zaatri}, A., \& {Corbard}, T. 2009, in Astronomical Society of the Pacific Conference Series, Vol. 416, Solar-Stellar Dynamos as Revealed by Helio- and Asteroseismology: GONG 2008/SOHO 21, ed. M.~{Dikpati}, T.~{Arentoft}, I.~{Gonz{\'a}lez Hern{\'a}ndez}, C.~{Lindsey}, \& F.~{Hill}, 99

\bibitem[{{Zhao} {et~al.}(2012){Zhao}, {Nagashima}, {Bogart}, {Kosovichev}, \& {Duvall}}]{zhao2012}
{Zhao}, J., {Nagashima}, K., {Bogart}, R.~S., {Kosovichev}, A.~G., \& {Duvall}, T.~L., J. 2012, \apjl, 749, L5, \dodoi{10.1088/2041-8205/749/1/L5}

\end{thebibliography}
\bibliographystyle{aasjournal}

\end{document}